# Surface Plasmons and Nonlocality – a Simple Model

Yu Luo[*], A. I. Fernandez-Dominguez, Aeneas Wiener, S. A. Maier & J. B. Pendry

*The Blackett Laboratory, Department of Physics, Imperial College London, London SW7 2AZ, UK*

[*]*Email address of the corresponding author: y.luo09@imperial.ac.uk*

## Abstract

Surface plasmons on metals can concentrate light into sub-nanometric volumes and on these near atomic length scales the electronic response at the metal interface is smeared out over a Thomas-Fermi screening length. This nonlocality is a barrier to a good understanding of atomic scale response to light and complicates the practical matter of computing the fields. In this letter, we present a local analogue model and show that spatial nonlocality can be represented by replacing the nonlocal metal with a composite material, comprising a thin dielectric layer on top of a local metal. This method not only makes possible the quantitative analysis of nonlocal effects in complex plasmonic phenomena with unprecedented simplicity and physical insight, but also offers great practical advantages in their numerical treatment.

Light incident on metallic nanoparticles excites surface plasmons, and these in turn can concentrate the light into sub-nanometre crevices or corners, giving rise to massive local energy density. This effect may be exploited to enhance spectroscopies or nonlinear phenomena [1, 2]. On the near-atomic length scale, the finite Fermi wavelength of electrons is a significant factor and any screening charge at an interface is smeared out over the Thomas-Fermi screening length. This nonlocal response contrasts with the usual assumption (valid on length scales >> 1nm) that the screening charge sits precisely at the interface, and limits the degree to which light can be concentrated [3, 4]. It also greatly complicates a theoretical treatment of sub-nanoscale light. First-principle methods [5, 6] describing the electron-ion dynamics in metals are available, where purely quantum mechanisms emerge naturally [7-9]. However, the enormous number of degrees of freedom involved prevents their application to realistic devices. This has given rise to an increasing interest in the development of semiclassical approaches which, retaining the physical insight and predictive value of classical electrodynamics, incorporate nonlocal effects into Maxwell's equations. For decades, the hydrodynamic model [10] has been extensively used to reflect the impact of nonlocality in the optical properties of tiny metal structures [11-22]. To describe the smearing of induced charges at the metal boundaries, this model requires the implementation of spatially-dispersive dielectric functions, which increases significantly the difficulty of the theoretical treatment, especially when dealing with realistic nanostructure geometries.

Here we show that what at first appears to be a complex effect can be represented by an extremely simple model: the effects of nonlocality can be reproduced to a high degree of precision by a model that replaces the nonlocal metallic surface with an effective local metallic surface coated with a dielectric layer.

Not only does this render computation a much simpler process, but also and of greater importance, the new model allows is to bring to bear on nonlocal problems all our intuitive understanding of local systems making for greater creativity and invention at the sub-nanometre scale.

It is well known that two particles are indistinguishable if they present the same scattering and extinction properties, both in the near and far field, and irrespective of the incident radiation. Inspired by this fact, we intend to find a general prescription for a virtual local structure which replaces a nonlocal plasmonic system and whose optical response can be easily described within the framework of conventional electromagnetics (EM). Thus, our starting point is the simplest geometry possible: a single metal-dielectric interface as the one depicted in Fig. 1 (a). Within the conventional nonlocal description, the reflection and transmission from this structure can be obtained through the hydrodynamic model [19]. In this model, the metal is characterized by a spatially-dispersive permittivity tensor of components $\varepsilon_\mathrm{T}(\omega) = \varepsilon_\infty - \omega_\mathrm{P}^2 / [\omega(\omega+i\gamma)]$ and $\varepsilon_\mathrm{L}(\mathbf{k},\omega) = \varepsilon_\infty - \omega_\mathrm{P}^2 / [\omega(\omega+i\gamma) - \beta^2 |\mathbf{k}|^2]$ for transverse and longitudinal EM fields, respectively. Here $\gamma$ and $\omega_\mathrm{P}$ denote the metal damping and plasma frequency, and the $\beta$ factor, proportional to the Fermi velocity, measures the degree of nonlocality. Owing to the spatially-dispersive character of $\varepsilon_\mathrm{L}(\mathbf{k},\omega)$, the excitation of longitudinal plasmon modes below $\omega_\mathrm{P}$ is possible. This greatly complicates the calculation, as compared with the local approximation. To avoid the implementation of a **k**-dependent permittivity, we next explore the possibility of mapping the nonlocal smearing of induced surface charges into a thin dielectric layer placed on top of the metal boundary [see Fig. 1(b)]. In what follows, we term the resulting system as the *local analogue model* (LAM).

Exploiting the fact that the excitation of longitudinal plasmons mainly affects the surface charge distribution across the metal boundary, our approach assumes that this nonlocal delocalization can be effectively described through the alteration of the material properties close to the metal/dielectric interface. In the LAM, EM fields are purely transverse, and the bulk metal permittivity is given by $\varepsilon_m = \varepsilon_T(\omega)$. We next look for the thickness and dielectric function of the cover layer so that the LAM yields the same transmission and reflection properties as the original nonlocal system. Note that the analogy must hold both in the near and far field, independently of the incident frequency and parallel wavevector. Detailed derivations provided in the supplementary material demonstrate that this requirement can be fulfilled as long as (i) the thickness, $\Delta d$, of the dielectric cover is much smaller than the metal skin depth; (ii) the dielectric constant of this layer, $\varepsilon_t$, is proportional to $\Delta d$ and obeys the following condition:

$$\frac{\varepsilon_t}{\Delta d} = \frac{\varepsilon_b \varepsilon_m q_L}{\varepsilon_m - \varepsilon_b}, \tag{1}$$

where $\varepsilon_b$ is the dielectric constant of the background, and $q_L = \sqrt{\omega_P^2/\varepsilon_\infty - \omega(\omega+i\gamma)}/\beta$ is the longitudinal plasmon normal wavevector, which is inversely proportional to the decaying length of the surface charges. Note that the small $\Delta d$ constraint can be relaxed by considering an anisotropic cover layer, whose tangential permittivity vanishes, while the normal component conforms to the $\Delta d$ dependence given by Eq. (1). To simplify our theoretical study, we give details of this possibility in the supplementary material, and consider here only the isotropic LAM in the following discussion. Without loss of generality, we assume that the metal is gold, with fitting parameters $\varepsilon_\infty = 1$, $\omega_P = 3.3$ eV, $\gamma = 0.165$ eV, and $\beta = 0.0036c$.

Figure 1 (c) displays the real and imaginary parts of $\varepsilon_t / \Delta d$ from zero to the bulk plasma frequency $\omega_p$. It is the aim of this work to demonstrate that Eq. (1) is universal and that the dielectric cover extracted from it describes accurately the emergence of nonlocal effects in a variety of plasmonic systems, provided that the metal thickness is larger than the Thomas-Fermi wavelength (~ 0.15 nm for gold [23]). Note that the dielectric layer is not uniquely defined, as only the ratio between its permittivity and thickness is fixed as Eq. (1). Thus, we can set $\Delta d$ as a constant and allow the permittivity to vary with frequency. This is equivalent to effectively modifying the material properties close to the metal boundary. Alternatively, we can assume $\varepsilon_t = 1$ and obtain a frequency-dependent thickness. This can be interpreted as a geometric shifting of the metal-dielectric interface. Our approach clarifies which of these two models mimics better the spatial nonlocality.

We explore the validity of the two interpretations above by investigating the dispersion characteristics of the surface plasmon polariton (SPP) mode supported by the metal-insulator-metal geometry in Fig. 2(a). The corresponding LAM geometry is depicted in Fig. 2(b). Both LAMs (one with an effective cover layer of a constant thickness, $\Delta d = 0.1$ nm, and one with a constant permittivity, $\varepsilon_t = 1$) are considered. Importantly, the cover layers in both cases are placed at the structural interfaces and extend into the metal regions, in such a way that the gap width remains the same as in the original geometry. Figure 2(c) shows the comparison of the SPP bands for a gap width $d = 1$ nm calculated using four different approaches. The cyan dashed line and green circles plot the local and nonlocal hydrodynamic results, respectively, where we observe the well-known spectral blue-shift due to nonlocality [19]. The predictions obtained from our LAMs are rendered in red solid (constant thickness) and grey dotted (constant permittivity) lines. On the one hand, the model assuming constant

$\Delta d$ agrees perfectly with the conventional nonlocal calculations over the whole frequency range. This implies that the modification to the material permittivity at the metal boundaries describes successfully the emergence of nonlocal effects in the geometry. On the other hand, the calculations based on $\varepsilon_t = 1$ yield a good comparison with nonlocal predictions only well below the surface plasmon frequency, and fail to reproduce the nonlocal blueshift of the SPP band. We can conclude that the interpretation of nonlocal effects as a geometric shifting of the metal boundaries is valid only at low frequencies, where the surface charge thickness (given by $1/q_L$) is much smaller than the metal skin depth.

The local description of spatial dispersion in metal permittivities simplifies drastically the theoretical treatment and offers physical insight into plasmonic phenomena in complex geometries. To illustrate this, we consider first a dimer of gold nanowires separated by a small gap, as shown in the inset of Fig. 3(b). Previous theoretical studies showed that this structure can be treated using a quasi-analytical transformation optics approach [24, 25]. However, once nonlocal effects are taken into account, the problem becomes much more complicated and only an approximate treatment is possible [26]. This is because $\varepsilon_L(\mathbf{k}, \omega)$ associated with the longitudinal field is not preserved under conformal transformations. Our LAM solves this problem as it eliminates the implementation of $\varepsilon_L(\mathbf{k}, \omega)$. Using the constant $\Delta d$ approach, the exact solution can be found through the manipulation of a tri-diagonal matrix. Using the $\varepsilon_t = 1$ assumption, the smearing of surface charges is simply mapped to the geometric modification, and we obtain a simple closed-form formula for the absorption cross-section and field enhancement. See detailed discussions of these two approaches in the supplementary material.

Figure 3(a) and (b) show the absorption and gap field enhancement spectra for gold nanowire dimers with $R = 10$ nm. The distance between the nanowires is set as 0.5 nm (results for other separations are provided in supplementary material). Four sets of calculations are performed, i.e. the local analytical calculations [25], nonlocal hydrodynamic simulations [22], LAM with $\Delta d = 0.1$ nm, and LAM with $\varepsilon_t = 1$. Calculations based on constant $\Delta d$ are again in excellent agreement with numerical simulations for all frequencies. The shifting boundary approach ($\varepsilon_t = 1$) also yields accurate predictions for the lowest resonance sustained by the structure, whereas at higher frequencies, this method leads to approximate spectra, presenting only slight deviations from the numerical results. In other words, the LAM evidences that the nonlocal effect here can be interpreted as adding a separation between the two nanoparticles. Thus, the field enhancement cannot be infinitely increased even for touching cylinders. This point is illustrated by Fig. 3(c) and (d), which show that, as recently reported experimentally [3], spatial nonlocality truncates the continuous redshift experienced by the plasmonic resonances, setting as well the limit for EM enhancements for separations above the quantum tunnelling regime [27-29].

Our approach does not only facilitate the description of nonlocality in plasmonic phenomena which can be tackled in an analytical fashion. It also makes possible a more efficient numerical treatment of these effects in nanoparticle geometries where such analytical description is not possible. Using the finite element method, the implementation of the nonlocal hydrodynamic model requires the simultaneous solution of both Maxwell's Equations and the transport equation for the nonlocal current [20-22]. In contrast, the LAM can be straightforwardly solved within any commercially available EM simulation platform. Apart from this evident simplification, our method is more convenient than the hydrodynamic treatment in terms of technical aspects such as the memory consumption and the calculation time. In the supplementary material, a comprehensive comparison is provided between both methods for the case of 3D gold nanoparticles of sizes ranging from 2 to 100 nm.

Figure 4 displays the numerical absorption (a) and the field enhancement (b) spectra for a three-dimensional gold conical dimer. The geometry of the dimer is sketched in the top right inset of panel (a). Note that all the geometric edges are chamfered with a 2 nm rounding radius, which is closed to the realistic situation. The left inset of Fig. 4(a) depicts the absorption spectra for a single isolated conical nanoparticle. In all cases, four different sets of data are shown. Local (conventional nonlocal) results are plotted in cyan dashed line (green circles). The predictions from the LAM with $\Delta d = 0.1$ and 0.01 nm are rendered in grey dotted and red solid lines, respectively. Our method is in excellent agreement with the hydrodynamic calculations, reproducing the blueshift experienced by the plasmonic resonances and the field enhancements at the gap centre with high accuracy. In order to further demonstrate the validity of our LAM in the near field, the insets of Fig. 4(b) display the electric field amplitude map within the gap region at resonance ($\omega = 0.35\omega_P$). The upper and lower insets correspond to our approach for $\Delta d = 0.1$ nm and the hydrodynamic simulation, respectively. The linear colour scale ranging from zero (blue) to 160 (red) is the same in both cases. We can observe that the LAM describes accurately the near-field characteristics of the nanostructure, failing only to reproduce the electric field profile within the metal regions. Here, we highlight that for the geometry considered in Fig. 4, our method was one order of magnitude more efficient (in terms of both time and memory) than the hydrodynamical treatment.

So far, we have demonstrated the validity of the LAM in predicting the surface modes below $\omega_P$. It can also be applied to study the bulk longitudinal plasmon resonances featured by small isolated nanoparticles at high frequencies [18, 30, 31]. Detailed studies (provided in the supplementary material) show that for nanoparticles with radius of curvature comparable to the Thomas-Fermi wavelength, Eq. (1) should be amended by

$$\frac{\varepsilon_t}{\Delta d} = \frac{\varepsilon_b \varepsilon_m q_L}{\varepsilon_m - \varepsilon_b} \frac{i_l'(q_L R)}{i_l(q_L R)} \qquad (2)$$

where $i_l(\cdot)$ and $i_l'(\cdot)$ are the modified spherical Bessel function of the first kind and its derivative, respectively. Note that, as expected, Eq. (2) reduces to Eq. (1) in the limit $R \gg 1/q_L$.

In the supplementary material (Fig. S7 and Fig. S8), we plot the absorption spectra for freestanding gold nanospheres with different radii. The results demonstrate that our LAM not only reproduces the nonlocal blueshift experienced by the low frequency surface mode, but also predicts the occurrence of bulk longitudinal plasmon resonances above $\omega_P$, which are absent in the local approximation.

Finally, let us remark that the local analogue strategy introduced in this work is not limited to the framework of the nonlocal hydrodynamic model. A similar methodology can be applied to more sophisticated mesoscopic treatments of quantum effects in plasmonics. In this context, electron spill-out [32, 33] may give rise to a negative dielectric function for the cover layer, which accounts for the quantum tunnelling effect. We leave detailed investigation of these effects for a later paper.

We have presented a local analogue model able to reflect the impact of nonlocality in the optical response of plasmonic structures. This model enables us to investigate realistic devices with Angstrom-sized features within the classical electro-dynamics framework. We have shown that the smearing of induced charges owing to nonlocality can be simply mapped into a geometric modification at low frequencies.

On the other hand, through the effective alteration of the material permittivity at the structural boundaries, nonlocal effects in arbitrary plasmonic systems can be accurately described over the whole frequency range. Finally, we have demonstrated that our approach not only provides a new understanding of plasmonic phenomena at the sub-nanometer scale, but also offers great practical advantages in their theoretical treatment, which will be of value in further studies of related phenomena, such as van der Waals interactions and nanoscale heat transfer.

**Acknowledgement**

This work was supported by the Leverhulme trust, the Gordon and Betty Moore Foundation, the AFOSR, and the EPSRC.

Reference

[1]  K. A. Willets, and R. P. Van Duyne, Annu. Rev. Phys. Chem. **58**, 267 (2007).
[2]  N. J. Halas, S. Lal, W. S. Chang, S. Link, and P. Nordlander, Chem. Rev. **111**, 3913 (2011).
[3]  C. Ciraci, R. T. Hill, J. J. Mock, Y. Urzhumov, A. I. Fernandez-Dominguez, S. A. Maier, J. B. Pendry, A. Chilkoti, and D. R. Smith, Science **337**, 1072 (2012).
[4]  J. A. Scholl, A. L. Koh, and J. A. Dionne, Nature **483**, 421 (2012).
[5]  G. Onida, L. Reining, and A. Rubio, Rev Mod Phys **74**, 601 (2002).
[6]  M. A. L. Marques, and E. K. U. Gross, Annu Rev Phys Chem **55**, 427 (2004).
[7]  J. Zuloaga, E. Prodan, and P. Nordlander, Nano Lett. **9**, 887 (2009).
[8]  J. Zuloaga, E. Prodan, and P. Nordlander, ACS Nano **4**, 5269 (2010).
[9]  D. C. Marinica, A. K. Kazansky, P. Nordlander, J. Aizpurua, and A. G. Borisov, Nano Lett. **12**, 1333 (2012).
[10] A. D. Boardman, *Electromagnetic surface modes* (Wiley, Chichester ; New York, 1982).
[11] A. D. Boardman, B. V. Paranjape, and Y. O. Nakamura, Phys Status Solidi B **75**, 347 (1976).
[12] G. S. Agarwal, Pattanay.Dn, and E. Wolf, Phys Rev B **10**, 1447 (1974).
[13] J. M. McMahon, S. K. Gray, and G. C. Schatz, Nano Lett. **10**, 3473 (2010).
[14] C. David, and F. J. Garcia-de-Abajo, J Phys Chem C **115**, 19470 (2011).
[15] A. I. Fernandez-Dominguez, A. Wiener, F. J. Garcia-Vidal, S. A. Maier, and J. B. Pendry, Phys. Rev. Lett. **108**, 023901 (2012).
[16] J. M. McMahon, S. K. Gray, and G. C. Schatz, Phys Rev Lett **103**, 097403 (2009).
[17] R. Ruppin, Phys Lett A **340**, 299 (2005).
[18] R. Ruppin, Opt Commun **190**, 205 (2001).
[19] A. Moreau, C. Ciraci, and D. R. Smith, Phys Rev B **87**, (2013).
[20] C. Ciraci, and D. R. Smith, Chemphyschem **14**, 1109 (2013).
[21] G. Toscano, S. Raza, A. P. Jauho, N. A. Mortensen, and M. Wubs, Optics Express **20**, 4176 (2012).
[22] A. Wiener, A. I. Fernandez-Dominguez, A. P. Horsfield, J. B. Pendry, and S. A. Maier, Nano Lett. **12**, 3308 (2012).
[23] N. W. Ashcroft, and N. D. Mermin, *Solid state physics* (Holt, New York,, 1976).
[24] R. C. Mcphedran, and W. T. Perrins, Appl. Phys. **24**, 311 (1981).
[25] A. Aubry, D. Y. Lei, S. A. Maier, and J. B. Pendry, Phys. Rev. Lett. **105**, 233901 (2010).
[26] A. I. Fernandez-Dominguez, P. Zhang, Y. Luo, S. A. Maier, F. J. Garcia-Vidal, and J. B. Pendry, Phys Rev B **86**, 241110 (2012).
[27] R. Esteban, A. G. Borisov, P. Nordlander, and J. Aizpurua, Nat. Commun. **3**, 825 (2012).
[28] K. J. Savage, M. M. Hawkeye, R. Esteban, A. G. Borisov, J. Aizpurua, and J. J. Baumberg, Nature **491**, 574 (2012).
[29] J. A. Scholl, A. Garcia-Etxarri, A. L. Koh, and J. A. Dionne, Nano Letters **13**, 564 (2013).
[30] M. Anderegg, Feuerbac.B, and B. Fitton, Physical Review Letters **27**, 1565 (1971).

[31]    I. Lindau, and P. O. Nilsson, Phys Scripta **3**, 87 (1971).
[32]    L. Stella, P. Zhang, F. J. García-Vidal, A. Rubio, and P. García-González, J. Phys. Chem. C **117**, 8941 (2013).
[33]    T. V. Teperik, P. Nordlander, J. Aizpurua, and A. G. Borisov, Physical Review Letters **110**, 263901 (2013).

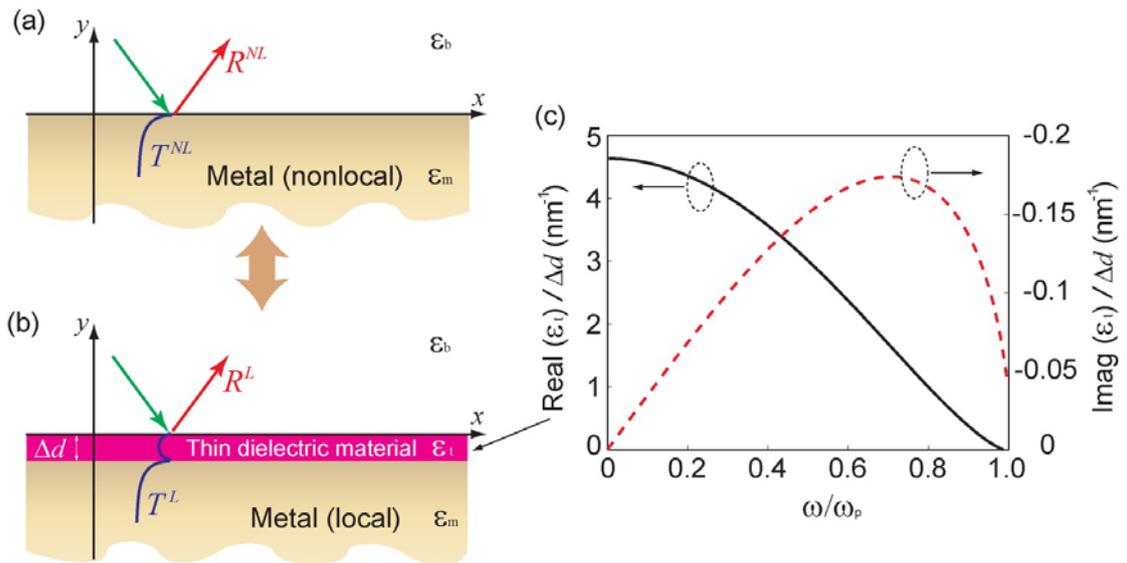

Figure 1 Schematics of a metal-dielectric interface within (a) the nonlocal description and (b) the LAM, where the surface charge smearing is mapped into a dielectric cover layer. Panel (c) plots the layer permittivity over its thickness as a function of the incident frequency for a gold Drude fitting. Note that the horizontal axis is normalized to the plasma frequency $\omega_P$.

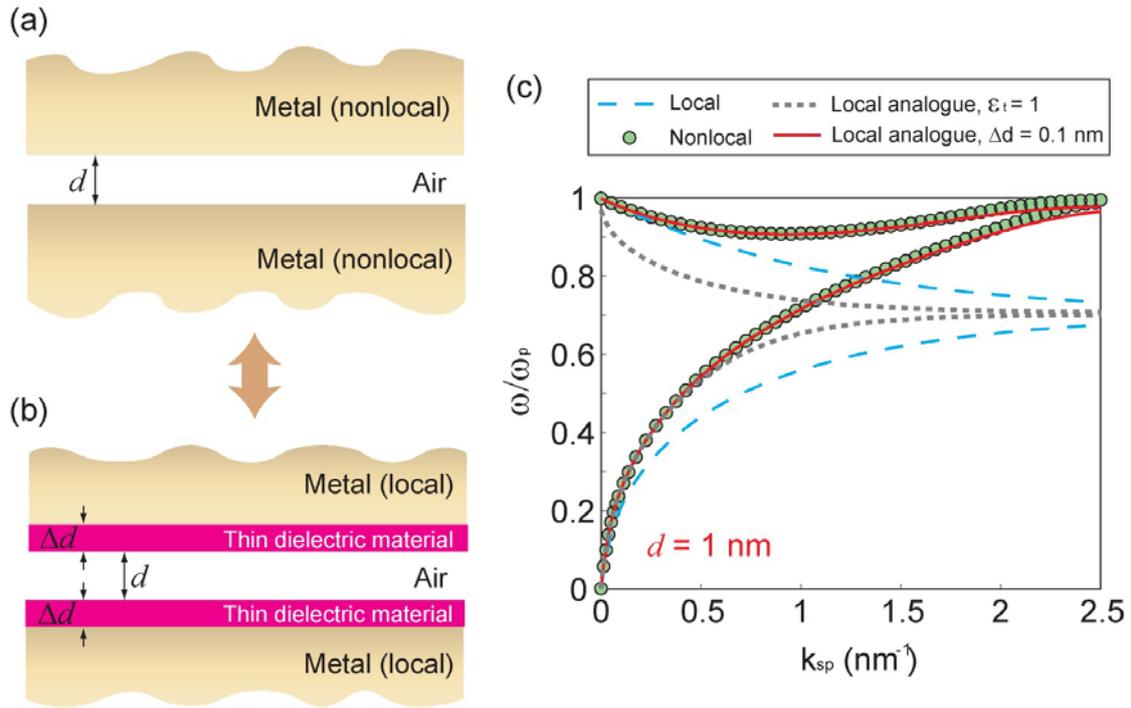

Figure 2 Schematics of a gold-air-gold geometry within (a) the nonlocal picture and (b) the LAM. Panel (c) shows the comparison of the SPP dispersion relation obtained from different models for a gap size $d = 1$ nm. Omitting the loss is not essential to our calculations, but gives a more critical test of our theory.

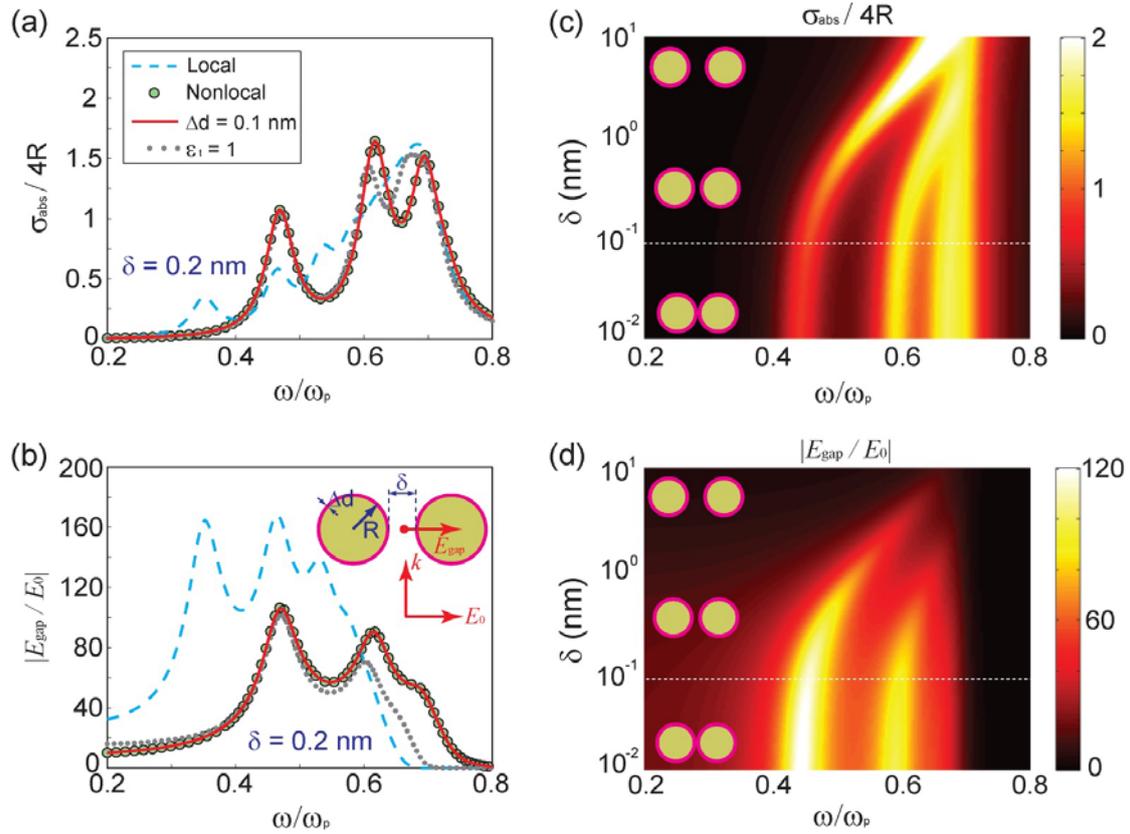

Figure 3 Optical response of a gold nanowire dimer within different treatments. (a) Absorption spectra for a pair of 10 nm radius gold nanowires separated by a $\delta = 0.2$ nm gap. (b) Electric field enhancement versus incident frequency evaluated at the gap centre for the same dimer geometry as in panel (a). (c) The absorption cross-section normalized to the dimer physical size obtained from Equation (2) as a function of $\omega/\omega_{\mathrm{P}}$ and $\delta$ (here $R = 10$ nm). (d) $E_{\mathrm{gap}}/E_0$ versus frequency and gap size. In all cases, the cyan dashed lines (green circles) correspond to local (nonlocal hydrodynamic) calculations. The red solid (grey dotted) lines plot the predictions from the LAM for constant $\Delta d$ (constant $\varepsilon_{\mathrm{t}}$).

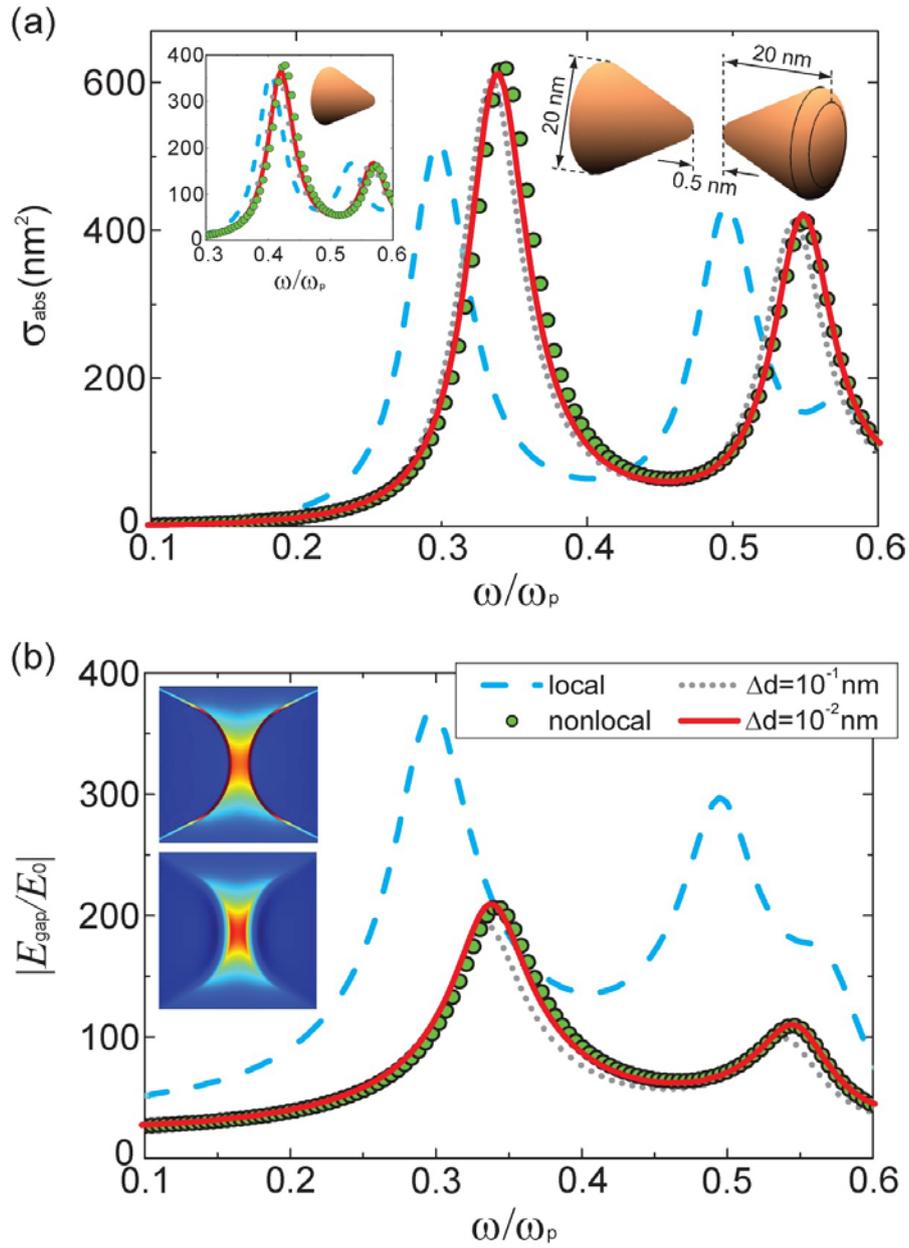

Figure 4 Numerical treatments of a 3D conical dimer within different models. (a) Absorption spectra for three-dimensional gold conical dimers with geometric parameters as shown in the top right inset of the panel. All the geometric edges are chamfered with a 2 nm rounding radius. The top left inset plots the absorption cross-section for a single conical nanoparticle. (b) Field enhancement spectra evaluated at the gap centre for the same structures as panel (a). The insets compare the electric field map at the gap region obtained numerically from the nonlocal hydrodynamic treatment (bottom) and the LAM with $\Delta d = 0.1$ nm (top).